# Social Interactive Media Tools and Knowledge Sharing: A Case Study

**Abstract**

*Purpose:* Social Media Tools (SMT) have provided new opportunities for libraries and librarians in the world. In academic libraries, we can use of them as a powerful tool for communication. This study is to determine the use of the social interactive media tools [Social Networking Tools (SNT), Social Bookmarking Tools (SBT), Image or Video Sharing Tools (IVShT), and Mashup Tools (MT)] in disseminating knowledge and information among librarians in the Limerick University, Ireland.

*Methodology:* The study was a descriptive survey. The research population included all librarians in Glucksman library. A questionnaire survey was done to collect data. Statistical software, SPSS16 was used at two levels (descriptive and inferential statistics) for data analyzing.

*Findings:* The findings show that the mean (out of 5.00) of using each of SMT in sharing knowledge by the librarians of Glucksman library is as the following: SNT (2.49), SBT (2.92), IVShT (2.99) and MT (2.5). It shows that most of their interaction related to share of image or video.

*Originality:* SMT provides an excellent platform for the exchange information between students, faculty members, and the librarians themselves. The Glucksman library at the University of Limerick is using this technology. This paper gives an example of how using these tools in the field of Library and Information Science in Ireland. The issues expressed could be beneficial for the development of library services in general and knowledge sharing among librarians in particular.

1. **Introduction**

In academic libraries, SMT counts as a new and powerful type of communication system that provides a good platform for knowledge sharing and it is an essential ground to attract the attention of both employees and patrons. Enemarka et al. (2012) identified how the structure of the communication network affects the efforts to coordinate or cooperate, and research suggests that adding connections to a network can improve the performance of groups when faced with such tasks. So, social networking can be a suitable environment for librarians' interaction to manage users' information needs in education and research.

This tool provides new and phenomenal facilities and opportunities for librarians which they can use from as a mean for the dynamics of their profession. For example, librarians with using SNT can create profiles and personal pages and the users also engage and network among them. In addition, librarians and library users with using SNT can be informed about all posts and links from other libraries and also express their views on the post by commenting on them. Generally, this tool is very important in order to update information and knowledge of the librarians and to cater to the informational needs of library users. Along with SNT, librarians can use from SBT to create subject heading and user- centered catalogs. In fact, its tagging and bookmarking capabilities make it possible for users to take part in the process of cataloging and creating of subject heading[1]. Image and video sharing tools are other SMT which librarians can use in order to introduce library resources and services, teach information literacy and provide a rich archive of photographs and films relating to conferences, seminars, and various lectures. They can also use MT to provide digital maps and use them as a resource for access to any SMT.

In fact, SMT is the most attractive tools that make it possible to exchange ideas among all peoples, especially librarians and create an appropriate context for the sharing of knowledge and online information.

The ultimate goal of SMT is creating an active and knowledge network community that individuals can exchange and share their valuable information.

Nowadays, many librarians are aware of the importance and application of these tools in their libraries and, as we see in many libraries in the world (national, public and academic) use SMT to provide services to visitors and to create an environment of mutual interaction in order to create an environment for sharing views, experiences, ideas and opinions with others. There are few libraries that do not benefit from these tools. The Glucksman Library, University of Limerick, Ireland, use the Web 2.0 tools and SMT on their website.

---

[1] The most specific word or phrase that describes the subject, or one of the subjects, of a work, selected from a list of preferred terms (controlled vocabulary) and assigned as an added entry in the bibliographic record to serve as an access point in the library catalog. A subject heading may be subdivided by the addition of subheadings (example: Libraries--History--20th century) (Reitz, 2013).

This study investigates the use of social interactive media tools in the process of knowledge sharing among librarians of Glucksman and thus will answer to following questions:

1. What is the extent of the use of SNT in the process of knowledge sharing among librarians of Glucksman?

2. What is the extent of the use of SBT in the process of knowledge sharing among librarians Glucksman?

3. What is the extent the Use of image and video sharing tools for sharing knowledge among librarians of Glucksman library?

4. What is the extent the use of integrated technology for knowledge sharing among librarians Glucksman library?

5. What difference is there among the use of various social interactive media and sharing of knowledge among librarians Glucksman according to the demographic profile (age, sex, degree, field and training)?

## 2. Literature Review

Thus far there has not been any research about the use of interactive SMT in the process of knowledge sharing among librarians, thus we tried to introduce some of the related research to the topic of our study.
 One of these researches about the study of knowledge sharing in social networking sites was done by Drula (2009). He conducted his research to look for users of social networks like Facebook, MySpace, LinkedIn and tagging sites, and ultimately concludes that among these social networks, MySpace and tagging sites didn't have many users but other sites like Facebook and LinkedIn had many users.

Van Zyl (2009) investigated the effectiveness of social networks in organizations. This research also aimed to educate IT, business decision makers, knowledge workers and librarians about the various applications, benefits and risks associated with social networking. Ultimately, it concluded that applying this type of web 2.0 tools in the organization will help people to help each other to engage in knowledge management.

Rotha and Cointet (2010) introduced a theoretical framework based on a social network and a socio-semantic network, i.e. an epistemic network featuring agents, concepts and links between agents and between agents and concepts. Adopting a relevant empirical protocol, they then described the joint dynamics of social and socio-semantic structures, at both macroscopic and microscopic scales, emphasizing the remarkable stability of these macroscopic properties in spite of a vivid local, agent-based network dynamics.

Anbari (2010) evaluated the specialized Farsi online social networks and its role in knowledge management and providing an appropriate model. The results show that performance of internal networks in satisfying the needs of users, encouraging them to participate in knowledge sharing, attracting trust and confidence of users, effectiveness in improving levels of specialized knowledge and its role in increasing the interaction of the user, is moderate and low.

Chaib (2010) investigated the use of SMT in the library's website of universities in Algeria. His results showed that level of using of SMT in libraries was low. Harinarayana and Raju (2010), aspects of their research using web 2.0 tools in the websites of academic libraries and concluded that most academic libraries in the dissemination of news and information about library use of RSS and Blog. While wiki, podcast and video tools allocated most less users in the area of services.

Ingebricson (2010) in a case study examined the impact of yammer technology in the process of knowledge sharing in a Multinational Consultancy Company. The results showed that Yammer technology and its facilities; create a new and effective communication channel between employees.

Kim and Abbas (2010) examine the functions of the web 2.0 in academic libraries, based on knowledge management perspective. Their findings show that the web 2.0, RSS tools and blog used very much in academic libraries and Tagging tools have been widely used by students.
Luyt, et al. (2010) in their study, evaluated and assessed the librarian's conception in the Singapore National Library about Wikipedia technology. Results show that all the librarians who were interviewed, are aware of the technology of Wikipedia.

Wahlroos (2010), in his thesis entitled "Social media as a form of organizational knowledge sharing: a case study on employee participation at Wartsila", investigated the role of SMT is in the sharing of knowledge. The results of his research showed that personal factors (using of this tool in personal life), organizational factors (activities of managers and coworkers and organizational guides) and technical factors such as technical skills in the use of SMT is effective in sharing of knowledge.

Mesrinejad in her Research (2011) investigated the possibility of humanitarian issues in Iranian academic libraries using web 2.0 tools and also the rate and extent of knowledge and attitude of staff towards using web 2.0 tools has been studied. The results showed that academic staff has been awareness about web 2.0 tools and except podcasts, all tools used to an acceptable level.

Asemi and Talkhabi (2012) in a research, investigated the level of awareness, usage and attitudes of graduate students of Sharif University about social interactive media web 2.0 and eventually concluded that among the seven groups of SMT in this study (including SNT, blogging tools, micro-Blogging tools, SBT, IVShT and video conferencing tools), wiki and micro-blogging are devoted maximum and minimum users to itself, respectively.

Wang and Wei (2011) in his study titled " knowledge sharing in wiki community: an experimental study" examined the role of wiki tools in knowledge sharing. Based on the results, wiki tools have been a positive effect on the sharing of knowledge among members of the research community.

In general, with regards to propounded studies, it seems that social interactive media tools have been effective in the process of sharing knowledge among the people such as librarians.

### 3. Web 2.0, Social media and Library

The term web 2.0 at first time was introduced by Dale Doyhgerty and Tim OReilly in April 2004. Web 2.0, at first has been just a claim, which alleged that the Internet will rise again, but when an OReilly media company, attended in the first web conference term of web was generalized and promised the online revolution (Levy, 2009). According to Miller (2006) web 2.0 is a term that is applied in relation to often associated with the increasing evolution of the World Wide Web. This evolution and transformation have been done from a set of web sites toward developing computer databases that serve to end-users of applied web programs.

Miller believes that web 2.0 technology doesn't mean that a new and completely different version of the web is created, but is a symbol of evolution and a composition of the complementary characteristics of the web 1.0, that increases scientific and social communications among web users, and its related technologies also facilitate communication, collaboration and cooperation among users.
Web 2.0, is participatory and this participation is often done from the end users, such as bloggers and others, therefore institutions and organizations will have equal opportunities to participate together.

 A user of web 2.0 is not only a content consumer but he is also producing a non-intensive mass process of content production. The idea of social media is also one of the results of web 2.0.
The term of social media was used for the first time in July 2006, by Chris Shipley (founder and director of global research group, guide wire). Shipley believes that social media is leader of future events for dialogues. This term also, is used by Tina Sharkey in 1997 to describe a type of the operator community of internet content (Reinhard, et al. 2012)

Damas (2004) also states that this term refers to any type of social media which people can use to interact with each other though in different places. Generally, the overall goal of social media is to establish an active community and knowledge network, by which users can exchange valuable information on the Web trough SMT. In other words, social media is a media created to increase accessibility and social interactions. Social media has changed human communication to a two-way interaction using web-based technology.

Librarians in libraries can use of library facilities and capabilities of web 2.0 and social media in order to facilitate information access and meet the information needs of their users and using social interactive media, select the best sources of information needed by the library community and save time of users. In fact, the increasing popularity of user-centric and user-friendly Internet services such as Facebook, MySpace, Flicker, YouTube, Wikipedia and other SMT have led to establishing library 2.0 .The term library 2.0, at first time was used by Michael Casey in 2005. He believes that the heart of library 2.0 is its user-center nature. In fact, this term indicates that the number of libraries using web 2.0 technologies and social media is increasing.

4. **Social media and knowledge sharing**

In general, several definitions of knowledge sharing are provided by different resources. For example, Bartol and Srivastava (2002) define knowledge sharing as a process during which the employees defuse their knowledge and information across their organization in a way through which people exchange their knowledge (implicit and explicit) jointly and create new common knowledge.

According to Wahlroos (2010), knowledge sharing encompasses two processes; knowledge is donating and collecting knowledge. Knowledge donating is the communication with others in the field of intellectual capital, while the emphasis of knowledge collecting is on cooperation and partnership with others in order to share the collective intellectual capital. Thus, when employees share their knowledge with others, it is the process of knowledge donating and when they discover the experiences of others, knowledge collecting process takes place. Here knowledge sharing means exchanging information, experiences, ideas and beliefs using knowledge donating and collecting (Figure 1).

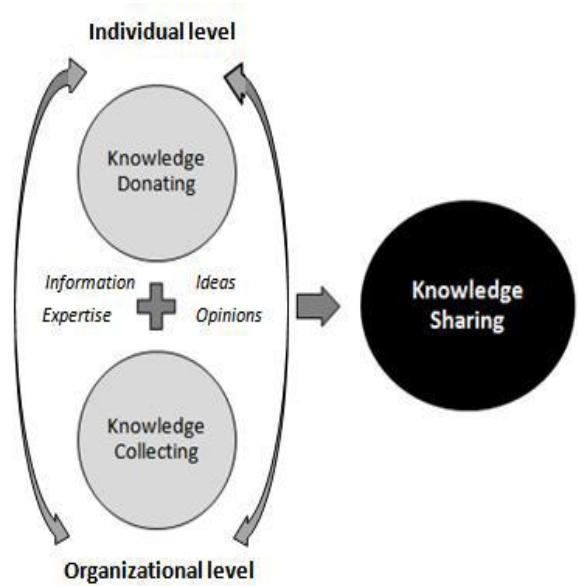

**Picture 1: Knowledge Sharing (Wahlroos, 2010)**

Currently the process of knowledge sharing with social media is introduced. Researchers believe that a shared web 2.0 has provided a common space for knowledge seekers and knowledge protectors and SMT is able to support both processes of knowledge donating and collecting, although SBT, image and video sharing tools, and RSS feeds have an active role in collecting and managing the knowledge. Tools such as blogs, wikis and social networks are considered the knowledge donating and among them, Wikis are the tools which applied to observe concepts (knowledge collecting), create and edit concepts (knowledge donating). Should be noted that the process of knowledge donating requires more time and efforts compared to knowledge collecting, because a person can quickly see the blog posts or wiki pages, but participation requires more time for SMT (Wahlroos, 2010)

This study also evaluates the use of social interactive media tools in the process of knowledge sharing among librarians at Glucksman, SMT is divided into four major instruments of social networks (including social networking sites, blog tools, micro-blogging tools and wiki tools), SBT (including social bookmarking sites, RSS tools and personalization tools), image and video sharing tools (including image and video sharing sites and video conferencing tools) and Mashup technology.

**Table1: Different Categories of SMT**
**(Along with tools for each group is displayed)**

| SMT | Sample |
|---|---|
| **SNT** | **Social networking sites** (Facebook , LinkedIn, My space, Fortuito, Nature network) **Blogging tools** (Scinceblog Blogger LiveJournal) **Micro-blogging tools** (Twitter, Yammer, Tumbler ) **Wiki tools** (Wikipedia,  ScienceWikia, Google Documents) |
| **SBT** | **Social bookmarking sites** (CiteULike, del.icio.us, digg, Connotea Flksonomy) **RSS tools** (Rss) **Personalization tools** (Icall, Igoogle, MyYahoo) |
| **IVShT** | **Image and video sharing sites** (Flickr, SlideShare, YouTube) **Video conferencing tools** (Instant Messenger, Skype, ooVoo, Google Video,  iTunes, Podcast) |
| **MT** | Mobile |

5. **Research Methodology**

The descriptive survey method was used in this study. This method is applied in studies that are seeking quantitative description of one or more aspects of affairs and thereby helps the researcher with obtaining to facts and comparing them, portraying result facts from quantitatively and qualitatively point of view.

In this study, the library method is used to collect and provide information about research background related to the topic and to identify all of the SMT, and a questionnaire survey was carried out in order to collect data for research purposes and to answer questions about research,. The statistical population of this research includes all the research librarians working in the Glucksman library which included 20 people and in five main units of the library namely the administrative unit, the information services unit, the unit of services to readers, the special sets unit and technical services unit.

Data analysis is presented in two levels of description and is based on statistics. In the descriptive level are used tables and charts for analyzing data. To examine equality means of two independent groups, two-sample T test and to determine the difference between means of several independent groups, ANOVA test are used. To investigate the relationship between two variables, Pierson correlation coefficient is applied.

## 6. Findings

To determine usage rate of social interactive media tools in knowledge sharing by Glucksman librarians, 33 questions were considered in total out of which 14 questions was related to SNT, 9 questions to SBT, 9 questions to image and video sharing tools, and 1 related to mash up technology. Since the comments of respondents has been collected in the form of Likert scale thus, for presenting the frequency Table based on Likert type, data obtained from averaging these questions were rounded to numbers 1,2,3,4,5 (Likert type).

**6-1. Using SNT in knowledge sharing among Glucksman librarians**

In order to investigate the effect of using SNT for knowledge sharing in Glucksman librarians, 14 questions were posed in the questionnaire related to social networking sites, blog tools, micro-blogging tools and wiki tools. These questions were designed to assess the effect of using social network tools in knowledge sharing in Glucksman librarians, mean of these 14 questions was considered as an overall view of the respondents. Table (2) shows the frequency and percentage of respondents about the effectiveness of using SNT in the sharing of knowledge is provided by librarians in this library.

**Table 2: Distribution of Respondents' Views about the Effectiveness of Using SNT in Knowledge Sharing**

| Using SNT in knowledge sharing | Very high | High | Medium | Low | Very low |
|---|---|---|---|---|---|
| F | 0 | 0 | 9 | 2 | 2 |
| F% | %0 | %0 | %14 | %3.1 | %3.1 |

Descriptive results were included frequency and mean for the effectiveness of using SNT in knowledge sharing are shown in in Table (3).

**Table 3: Frequency and Mean for the Effectiveness of Using SNT in Knowledge Sharing**

| Variable | f | Mean |
|---|---|---|
| The effectiveness of using SNT in knowledge sharing | 13 | 2.49 |

Also, to evaluate the effectiveness of each social network tools in knowledge sharing in Glucksman librarians, descriptive results include the frequency and mean is presented in Table (4).

**Table 4: Descriptive Results Include the Frequency and Mean for the Effectiveness of SNT in Knowledge Sharing**

| Variable | f | Mean |
|---|---|---|
| The effectiveness of using social networking sites in knowledge sharing | 13 | 2.38 |
| The effectiveness of using blogging tools in knowledge sharing | 13 | 2.36 |
| The effectiveness of using micro-blogging tools in knowledge sharing | 13 | 1.77 |
| The effectiveness of using wiki tools in knowledge sharing | 13 | 2.46 |

Accordance with the Table (3) SNT have the lowest use mean in knowledge sharing among Glucksman librarians, also, according to the Table (4) The highest average use of SNT in the study population is related to wiki tools (2.46) and the lowest average use is micro-blogging tools (1.77). The findings of this part of the research are consistent with findings of Asemi and Talkhabi (2012). This researcher in his study has investigated the awareness, using rate and attitudes of University graduate students to social interactive media, and ultimately concluded social network and micro-blogging tools haven't been so beneficial in the research cycle of students of Sharif University even though these students have used wiki tools well. Also according to findings of Wang and Wei (2011) who have investigated knowledge sharing across wiki community, wiki tools had a positive impact in knowledge sharing among surveyed community members.

It is found that among social networking sites, Facebook (3.54) and Fortuito (1.62) have respectively the highest and lowest mean in knowledge sharing. Also, among the blogging tools, highest and lowest mean is respectively related to blogger (3.77) and science blog (1.54). The results of this part of the study are consistent with findings of Drula (2009), who his article has investigated knowledge sharing in social networks and concluded Facebook and LinkedIn have the highest use and MySpace, the lowest use in knowledge sharing among Romanian social networks.

Furthermore, between micro-blogging tools, Twitter has the highest mean (2.39) while Yammer and Tumbler (1.15) had the lowest use average among Glucksman librarians, this finding is inconsistent with results of Ingebricson (2010) who has investigated the role and effectiveness of Yammer technology in knowledge sharing in employees of a multi-national company and concluded that the majority of them has used Yammer technology in their knowledge sharing and considered it as a new communication channel.

Between wiki tools, the highest and lowest use average is respectively related to Wikipedia (3.08) and science Wikia (1.31). According to findings of Luyt (2010) who investigated the perception level of librarians of librarians of the Singapore national library about Wikipedia tools, the majorities of them were aware of Wikipedia and used it.

**6-2. Using SBT in knowledge sharing among Glucksman librarians**

9 questions were posed in order to investigate the use of SBT among Glucksman librarians which were related to social bookmarking sites, RSS tools, and personalization tools respectively. To obtain the overall view of respondents about the effectiveness of using social bookmarking in knowledge sharing among Glucksman librarians, mean of these 9 questions was considered as an overall view of respondents. Frequency and frequency percent of respondents' views about the effectiveness of using bookmarking tools in knowledge sharing among Glucksman librarians, are presented in Table (5)

**Table 5: Frequency Distribution of Respondents' Views about the Effectiveness of Using SBT in Knowledge Sharing**

| Using SBT in knowledge sharing | Very high | High | Medium | Low | Very low |
|---|---|---|---|---|---|
| F | 0 | 4 | 5 | 4 | 0 |
| F% | %0 | %6.2 | %7.8 | %6.2 | %0 |

Descriptive results include the frequency and mean of the effectiveness of SBT in knowledge sharing are provided in Table (6).

**Table 6: Frequency and Mean of Effectiveness of SBT in Knowledge Sharing**

| Variable | f | Mean |
|---|---|---|
| Effectiveness of social bookmarking in knowledge sharing | 13 | 2.92 |

Also descriptive results include frequency and mean of effectiveness of SBT in knowledge sharing in this library librarians are provided in Table (7).

**Table 7: Frequency and Mean of Effectiveness of SBT in Knowledge Sharing**

| Variable | f | Mean |
|---|---|---|
| The effectiveness of social bookmarking sites in knowledge sharing | 13 | 2.55 |
| Effectiveness of RSS tools in knowledge sharing | 13 | 3.15 |
| Effectiveness of personalization tools in knowledge sharing | 13 | 3.05 |

According to Table (6) use of SBT among Glucksman librarians is near to middle, this result is consistent with the results of Asemi and Talkhabi (2012), where in his own research he has concluded that the use of SBT in the research cycle of Sharif University students are not so significant and as a result, using this tool hadn't been very useful for doing research.

Also, according to Table (7) The highest and lowest average in using of SBT among Glucksman library are related to RSS (3.15) and social bookmarking sites (2.55) respectively. In addition, among social bookmarking sites, Delicious site (3.23) and Connotea site (2.77) the highest and lowest average in knowledge sharing among Glucksman librarians have respectively and between personalization tools, the highest and the lowest average among this group of librarians is respectively related to call (3.23) and IGoogle (2.92).

The results of this part of the research are partly consistent with findings of Kim and Abbas (2010). Kim and Abbas in their research have investigated the functions of the web 2.0 based on knowledge management perspective in academic libraries and students and concluded that among the tools of web 2.0, RSS is used to a large extent in academic libraries and tagging tools are more common among students of this University. In addition, findings of Asemi and Talkhabi (2012) showed that the use of social bookmarking sites hasn't been beneficial in the research cycle of graduate students of Sharif University.

**6-3. using IVShT in knowledge sharing among Glucksman librarians**
To investigate the effect of image and video sharing tools in knowledge sharing among librarians of Glucksman library, 9 questions were raised in the questionnaire that related to image and video sharing sites and video conferencing tools. To achieve the overall result of the comments from respondents about the effectiveness of image and video sharing tools in knowledge sharing among Glucksman librarians, mean of these 9 questions were considered as the overall opinion of respondents. Frequency and frequency percentage of respondents' views about the effectiveness of the image and video sharing tools in knowledge sharing by librarians of this library are provided in Table (8).

**Table 8: Distribution of Respondents View about the Effectiveness of IVShT in Knowledge Sharing**

| Effectiveness of image and video sharing tools in knowledge sharing | Very high | High | Medium | Low | Very low |
|---|---|---|---|---|---|
| F | 0 | 3 | 7 | 3 | 0 |
| F% | %0 | %4.7 | %10.9 | %4.7 | %0 |

Descriptive results include frequency and mean of effectiveness of image and video sharing tools in knowledge sharing are provided in Table (9).

**Table 9: Frequency and Average of Effectiveness of Image and Video Sharing Tools in Knowledge Sharing**

| Variable | f | Mean |
|---|---|---|
| Effectiveness IVShT in knowledge sharing | 13 | 2.99 |

Also, descriptive results consisting frequency and average of effectiveness of each social tool for image and video sharing tools in knowledge sharing are presented in Table (10).

**Table 10: Frequency and Average of Effectiveness of IVShT in Knowledge Sharing**

| Variable | f | Mean |
|---|---|---|
| Effectiveness of video conferencing tools in knowledge sharing | 13 | 2.62 |
| Effectiveness of image and video sharing sites in knowledge sharing | 13 | 3.37 |

According to Table (9), image and video sharing tools have the highest average of use in the process of knowledge sharing among Glucksman librarians. This finding is inconsistent with results of Asemi and Talkhabi (2012) research. In his research has concluded that the use of this group of tools hasn't been so beneficial in doing investigative works among students of Sharif University. Also, based on findings of this part of research among image and video sharing sites to the highest and lowest use in librarians of this library, was related to slide sharing (3.54) and Flicker (3.23) respectively and the highest and the lowest use average of video conferencing tools in the study population, are respectively related to avoid software (3.31) and iTunes and podcasting technology (1.38).

The results of this part of the study are partly consistent with results of Chaib (2010) research; he evaluates the use of SMT in the web sites of Algiers libraries and ultimately concluded that in many websites of study libraries haven't been used Flickr and YouTube. Also, based on findings from research and Harinarayana and Raju (2010) use of podcasting technology in the libraries under study is very weak.

**6-4. Using the of MT in Knowledge Sharing among Glucksman Librarians**

Frequency and percent of frequency of respondents' views about the effect of MT in knowledge sharing which provided by librarians of this library, is presented in Table (11).

**Table 11: Distribution of Respondents' Views about the Effect of MT in Knowledge Sharing**

| Effect of Mashup technology in knowledge sharing | Very high | High | Medium | Low | Very low |
|---|---|---|---|---|---|
| F | 2 | 2 | 0 | 8 | 1 |
| F% | %3.1 | %3.1 | %0 | %12.5 | %1.6 |

Frequency and average of effectiveness of a Mashup technology in knowledge sharing are presented in Table (12).

**Table 12: Frequency and Average of Effectiveness of MT in Knowledge Sharing**

| Variable | f | Mean |
|---|---|---|
| Effectiveness of Mashup technology in Knowledge sharing | 13 | 2.5 |

**6-5. Relationship between demographic characteristics and use of SMT in knowledge sharing among Glucksman librarians**

Two-sample T-test was used to investigate the differences between male and female librarians of Glucksman library in use of SMT in the knowledge sharing.

Table (13) descriptive results include frequency and average for use of SMT is given to both sexes. T-test results in Table (14) are presented.

**Table 13: Frequency and Average for Use of SMT by Sex Segregation**

| Variable | Sex | f | Mean |
|---|---|---|---|
| Use of SMT | Male | 4 | 2.42 |
| | Female | 9 | 2.86 |

**Table 14: Table of T-test to Evaluate Differences between Male and Female Librarians of Glucksman Library in Use of SMT in Knowledge Sharing**

| T | df | P |
|---|---|---|
| 1.09 | 11 | 0.299 |

According to the Table (14) because the probable amount is equal to 0/299 and larger than 0/05 , the alleged differences between male and female librarians of Limerick University in terms of use of SMT in knowledge sharing is rejected at a significance level of 0/05.
The correlation coefficient was used to investigate the relationship between age and use of SMT. Results are presented in Table (15).

**Table 15: Pearson's Correlation Test for Determining the Relationship between Age and Use of SMT in Knowledge Sharing among Glucksman Librarians**

| p | Pearson's correlation coef. |
|---|---|
| 0.612 | 0/026 |

Given that probability amount is equal to 0.026 and smaller than 0/05, therefore, the claim that there is a lack of connection between age and use of SMT will be rejected at a significance level of 0/05, and because the correlation value is equal to 0.612, it can be said that there is a direct relationship between increased age and use of SMT.

Two-sample T-test was used to investigate whether use of SMT in knowledge sharing by librarians with Bachelor degree, librarians with Master degree and Ph.D. degree is the same. Descriptive results include frequency and the mean for the use of SMT by degree separation is given Table (16). T-test results are presented in Table (17).

**Table 16: Frequency and the Mean for the Use of SMT by Degree Separation**

| Variable | Education degree | f | Mean |
|---|---|---|---|
| Use of SMT | Bachelor | 8 | 2.88 |
| | Master's and Ph.D. | 5 | 2.48 |

**Table 17: T-test Table to Evaluate the Differences between the Use of SMT in Knowledge Sharing in Glucksman Librarians, by Education Degree**

| T | df | p |
|---|---|---|
| 1.062 | 11 | 0/311 |

According to Table (17) because probability amount is equal to 0.311 and larger than 0/05 thus, the alleged differences between bachelor and master librarians in the use of SMT - web 2.0 in knowledge sharing is rejected at significance level 0/05.
Two sample T-test was used to investigate whether the use of SMT to share the knowledge of libraries with librarianship and librarian with non-librarianship knowledge is the same whether the. Descriptive results include the frequency and mean of use of SMT variable are given to discipline separation in Table (18). T-test results are presented in Table (19).

**Table 18: The Frequency and Mean of Use of SMT Variable for Each Discipline**

| Variable | Discipline | f | Mean |
|---|---|---|---|
| Use of SMT | Librarianship | 5 | 2.43 |
| | Non-librarianship | 8 | 2.91 |

**Table 19: T-test Table to Evaluate Differences between the Use of SMT in Knowledge Sharing by Glucksman Librarians, based on Discipline**

| T | df | P |
|---|---|---|
| -1.295 | 11 | 0.222 |

According to Table (19) because probability amount is equal to 0/222 and is larger than 0/05, therefore, claimed differences between using social media in sharing the knowledge of librarians in the librarians librarianship and non-librarianship is rejected at significant level of 0/05.

ANOVA test was used to investigate whether the use of SMT interactive in knowledge sharing among librarians at the Glucksman librarians, with different training courses is the same. Descriptive results include the frequency and mean for the use of SMT in sharing the knowledge of librarians of the University of Limerick is variable by separating training courses related to the article's objective is provided in Table (20). Results of ANOVA test are presented in Table (21).

**Table 20: Frequency and Mean for the Use of SMT in Sharing the Knowledge of Glucksman by Separating Training Courses**

| Variable | Training course | Frequency | Mean |
|---|---|---|---|
| Use of SMT in sharing the knowledge of Glucksman librarians | MCSD | - | |
| | CCNA | - | |
| | Web Design | 3 | 2.52 |
| | +Network | - | |
| | ICDL | 3 | 2.52 |

**Table 21: Table of ANOVA Test for Investigating the Differences in the Use of SMT the Knowledge of Glucksman Librarians based on Training Courses**

| | Freedom degree | Sum of squares | Mean of squares | F | P |
|---|---|---|---|---|---|
| Between group | 2 | 0.484 | 0.242 | 0.488 | 0.628 |
| Within a group | 10 | 4.954 | 0.495 | | |
| Total | 12 | 5.437 | | | |

According to Table (21) because probability amount is equal to 0.628 and larger than 0/05, thus, the claimed difference in using SMT in sharing the knowledge of Glucksman librarians based on the spent training courses associated with the article's objective is rejected at significant level of 0/05. Simply put, the amount of use of SMT by Limerick University librarians in knowledge sharing is not to the type of training undergone by them.

According to Tables (14), (17), (19) and (21) there are no significant relationships between sexes, degree, discipline and training courses related to the usage of SMT by the librarian in knowledge sharing. This indicates that these factors haven't any effect on the rate of using SMT tools in knowledge sharing. However, based on the results of Table (15) there is a direct but weak relation between age and use of SMT in knowledge sharing among librarians in this university, so that, by increasing the age, use of SMT - interaction is increased.

**6-6. The overall goal of research: determining the use rate of SMT in knowledge sharing among Glucksman librarians**

As we mentioned previously, to investigate the effect of SMT in knowledge sharing among Glucksman librarians, 33 questions were posed in the questionnaire. The respondents' views were collected in the Likert scale than the average of these 33 questions were considered as the overall view of respondents. Frequency and frequency percentage of the respondents' views about the effectiveness of using SMT tools in sharing the knowledge of this library librarians is provided in Table (22).

**Table 22: Frequency Distribution Of Respondents' Views about the Effectiveness of Using SMT in Knowledge Sharing**

| The effectiveness of using SMT in knowledge sharing | Very high | High | Medium | Low | Very low |
|---|---|---|---|---|---|
| F | 0 | 2 | 6 | 5 | 0 |
| F% | %0 | %3.1 | %9.4 | %7.8 | %0 |

Descriptive results include the frequency and the mean for efficacy of SMT in the sharing of knowledge" in knowledge sharing has been provided in Table (23).

**Table 23: Frequency and the Mean for Efficacy of SMT in Knowledge Sharing**

| Variable | F | Mean |
|---|---|---|
| Efficacy of SMT in knowledge sharing | 13 | 2.73 |

7. **Conclusion**

In accordance to the findings of the research, the possible factors that haven't been ineffective in the results of this research will be discussed. As mentioned earlier, the highest and lowest average use of SNT in the process of knowledge sharing tools are related to wiki and micro-blogging tools. The probable reason can be attributed to the existence of these tools. As mentioned in the literature, wiki invention and micro-blogging invention refer to the period from 2001 to 2006. Consequently, wiki technology that is older than the micro-blogging technology, has been known earlier and accepted more than micro-blogging tools and thus is also used by people at a higher rate. Another factor t which leads to greater use of wiki tools can be attributed to presentation the content of Wikipedia in Google search results. As Luyt, et al (2010) findings show that Wikipedia was seen as a convenient information application as it is indexed by Google and appears among the top hits in many Google searches. This factor most likely has led to greater use of the blog tools in this university librarian. In fact, the reputation of a tool is important to generate more usage and more knowledge of a tool. Also, this research shows that librarians use more of Facebook (among the social networking sites), Blogger (among the blog tools), Twitter (among the micro-blogging tools), Twitter, and Wikipedia (among the wiki tools) in share of knowledge. These tools are more popular than the other tools. Perhaps the lack of use or less use of the My Space, Fortuito, Nature Network, Science Blogs, LiveJournal, Yammer, Tumbler, Science Wikia also be attributed to this factor.

In addition the highest use an average of SBT among this university librarian related to RSS technology and the lowest relates to social bookmarking sites. This may be due to, use directly of RSS technology in the library website. Another important factor that likely has led to greater use of RSS technology among librarians of the university library can be attributed to more reputation of this compare with SBT. Nowadays there are rare sites which doesn't use of RSS technology. More attention to this technology can lead to increased peoples' attention to this tool and increased use it. Consequently increased use of this tool to communicate with relatives and friends as a possible factor leads to greater use of this technology in the workplace and in connection with colleagues and greater use of the image and video sharing tools in comparing with other social interactive media tools can be attributed to this factor. Here this can be concluded that previous knowledge of an instrument, measure of acceptance, and greater use of a tool among members of a community, are important factors in the increasing use of that tool in different organizations and among members of these organizations.

And finally, findings showed that use of social media interactive tools among Glucksman librarians is lower than average. It requires more attention from the Glucksman librarians towards this tool. It is noteworthy that the librarians, other staff, and students can participate in a 12-week online, interactive, self-directed training program called 'the 23 things', run by the library. This course is a web 2.0 learning experience for the users. The purpose of this program on one hand is to increase awareness of employees, teachers and students about these tools in order to better perform in scientific activities and assist in the knowledge sharing among them and on the other hand to encourage library staff to use these tools more and to overcome the fear of this technology. This is a big step toward more use of this tool by librarians and library users of the Glucksman library of University of Limerick - Ireland.

## ACKNOWLEDGEMENT

We are very grateful to the people who have helped researchers in collecting data: Sinead.Keogh, Anne.McMahon; Aoife.Geraghty; Ciara.McCaffrey; Cora.Gleeson; Donna.ODoibhlin; Justine.Bennett; Ken.Bergin; Liz.Dore; Liz.O'Sullivan; Mary.Dundon; micheal.ohaodha; Michelle.Breen; Michelle.Dalton; Pattie.Punch; Peter.Reilly, and Gobnait.ORiordan, Glucksman Library, University of Limerick, Ireland.